\documentclass[12pt]{article}
\usepackage{epsfig,multicol}
\usepackage[english]{babel}
\usepackage{graphicx}
\usepackage{rotating}
\usepackage{amsmath}
\usepackage{amssymb}
\usepackage{flafter}

\def\brmu{{\rm BR}(\mu\rightarrow e\gamma)}
\def\brtm{{\rm BR}(\tau\rightarrow \mu\gamma)}
\def\mue{\mu\rightarrow e\gamma}
\def\tmu{\tau\rightarrow \mu\gamma}
\def\ie{{\em i.e.}}
\newcommand{\be}{\begin{equation}}
\newcommand{\ee}{\end{equation}}

\begin{document}

\begin{titlepage}

\begin{flushright}
DFPD-04/TH-01\\
SISSA 4/2004/EP 
\end{flushright}

\begin{center}

\vspace*{2cm}
{\LARGE Lepton Flavor Violation, Neutralino Dark Matter\\[0.3cm] and the Reach of the LHC}\\
\vspace*{1.5cm}
{\bf\large A.~Masiero$^1$, S.~Profumo$^2$, S.~K.~Vempati$^1$ and C.~E.~Yaguna$^2$}\\[0.7cm]
{\em $^1$ Dipartimento di Fisica `G.~Galilei', Universit\`a di Padova,
and Istituto Nazionale di Fisica Nucleare, Sezione di Padova, 
Via Marzolo 8, I-35131, Padova, Italy}\\[0.3cm] 
 {\em $^2$ Scuola Internazionale Superiore di Studi Avanzati,
Via Beirut 2-4, I-34014 Trieste, Italy
and Istituto Nazionale di Fisica Nucleare, Sezione di Trieste, 
I-34014 Trieste, Italy}\\[0.7cm] 
{\em E-mail:} {\tt antonio.masiero@pd.infn.it, profumo@sissa.it,\\ 
sudhir.vempati@pd.infn.it, yaguna@sissa.it}

\vspace*{0.8cm}

\begin{abstract}
\noindent We revisit the phenomenology of the Constrained MSSM with right-handed neutrinos (CMSSMRN). A supersymmetric seesaw mechanism, generating neutrino masses and sizable lepton flavour violating 
(LFV) entries is assumed to be operative.  In this scheme, 
we study the complementarity between the `observable ranges' of various paths leading 
to the possible discovery of low energy SUSY: the reach of the Cern Large Hadron Collider (LHC), 
the quest for neutralino dark matter signals and indirect searches through LFV processes.
Within the regions of the CMSSMRN parameter space compatible with all 
cosmo-phenomenological requirements, those which are expected to be probed at the LHC 
will be typically also accessible to upcoming LFV experiments. Moreover, parameter 
space portions featuring a heavy SUSY particle spectrum could be well beyond LHC reach 
while leaving LFV searches as the only key to get a glimpse on SUSY. 
\end{abstract}  
\end{center}

\end{titlepage}

\section{Introduction}\label{sec:introduction}
This is going to be the decade where we should be able to establish
whether low energy supersymmetry (SUSY) exists or not. We have three
main roads to gain access to SUSY: (a) Direct SUSY searches at hadron
colliders (b) Indirect SUSY searches in rare FCNC (Flavour Changing 
Neutral Currents) and/or CP violating  processes (c) Direct and indirect
SUSY Dark Matter (DM) searches. Given that low energy supersymmetry is 
rather a {\em framework} than a {\em predictive model}, one has to assume
an explicit realisation of SUSY breaking in order to make predictions
concerning the above mentioned three roads to SUSY. The Constrained Minimal 
SUSY Standard Model (CMSSM) represents the `prototype' of low
energy supergravity having the minimal number of free parameters 
and being successful in not departing too violently from the Standard Model 
(SM) predictions. In view of the extraordinary agreement of the SM with all
the experimental data on the above (a) and (b) points, it is of interest to
consider models with features approaching those of the CMSSM. 

The CMSSM, in its original formulation, is however not complete, and 
it has to be supplemented with some mechanism to generate non-zero neutrino 
masses and mixings. The supersymmetric seesaw mechanism represents a very attractive
candidate to accomplish this task. Hence we are led to study a CMSSM with 
right handed neutrinos (CMSSMRN).  We consider this as the `minimal' low-energy
supergravity-inspired setting with massive neutrinos.  Compared to the CMSSM, the main novelty 
in the CMSSMRN is the presence of lepton flavour violation (LFV) \cite{fbam}. Information 
from neutrino masses is nonetheless not sufficient to determine all 
the seesaw parameters \cite{casasibarra}, which are crucial to compute the 
relevant LFV rates. To remedy this, either a top-down approach with specific 
SUSY-GUT models and/or flavour symmetries \cite{lfv2,ue3importance,lopsidedlfv,oscar} 
or a bottom-up approach with specific parameterisations of low energy unknowns have been 
adopted in the literature \cite{lfv1,petcovyag1,petcovyag2}.
In the present work, we follow a hybrid approach making the minimal assumptions required to 
successfully predict lepton flavour violation. Assuming one of the 
neutrino Yukawa couplings to be as large as the top Yukawa, and varying the other
relevant seesaw parameters, we move from situations where LFV rates are significantly large, and already constrain the CMSSMRN parameter space (what we call the ``{\em best-case}'' scenario) to cases where a significant experimental improvement would be needed to detect flavour violation in the lepton sector (``{\em worst-case}'' scenario). 

In the light of this setting, we can now revisit the above three roads leading to 
SUSY discovery. Namely, we wish to conduct a quantitative analysis taking into 
consideration (a) direct search limits for SUSY at LEP and hadron colliders 
(b) indirect searches in rare FCNC processes, particularly concentrating on LFV,
and finally (c) the neutralino relic density, which should respect the upper bound
on the cold dark matter (CDM) content of the universe. We outline a strong 
complementarity between the three mentioned roads for SUSY discovery. We generically find, in 
the ``best case'' scenario, that the inclusion of the LFV constraints significantly 
restricts the already ``Constrained'' MSSM parameter space. In particular, in 
the most favorable cases, if stau coannihilations are responsible for the suppression of 
the neutralino relic density, SUSY might `appear' in LFV experiments even before the 
advent of the LHC, which will thoroughly probe these regions. In the other two parameter 
space areas allowed by the stringent WMAP upper bound \cite{wmap} on the CDM content, namely the
A-pole funnel and focus point regions, LFV sensitivity is found to be usually larger than the LHC reach,
provided the unknown leptonic mixing angle $U_{e3}$ is not too small. 
Indeed, given that, particularly in the focus point
region, SUSY masses are quite heavy, it might well be that LFV constitutes
the only tool at our disposal to get SUSY signals. Therefore, in the ``best-case'' situation, 
the inclusion of massive neutrinos has {\em critical} implications on the discovery paths for SUSY.  
On the other hand, in the ``worst-case'' scenario, LFV experimental studies will at best only be {\em complementary} to LHC direct SUSY searches. 
In this case, non-zero neutrino masses do not have a significant impact on
the CMSSMRN phenomenology, which would be almost indistinguishable from that of the CMSSM. 

Some interesting analyses of the impact of LFV on 
the CMSSM parameter space already exist: for instance, in Ref.~\cite{campbell} the correlation between
DM and LFV was considered in a bottom-up approach. Recently, 
 in Ref.~\cite{kingblazekfull} the connections between rare processes 
and seesaw generated LFV in CMSSM have been studied. Our analysis simultaneously takes into 
account high energy collider physics and low energy flavour constraints as well as the latest information on the DM content of the universe. Moreover, the
hybrid approach on the seesaw implementation that we adopt
here differs from what was done in previous works. We detail our 
results in the following Sections. 

In the next Section, we review the CMSSMRN model and present our assumptions. 
We also briefly review the present status of `discovery limits' of SUSY within the three 
discovery methodologies.  In Sections 3 and 4, we provide full details on our numerical 
results, mostly concentrating
on the ``best case'' scenario. In Section 5, we stress the role of $U_{e3}$ in 
connection with direct and indirect SUSY searches. After a brief
summary of the results, we conclude with an outlook.  

\section{The CMSSMRN and LFV}\label{sec:model}
The CMSSM is based on minimal supergravity with universal soft terms at the
GUT scale, $M_{GUT}\approx 2\cdot 10^{16}$ GeV. The SUSY particles spectrum at the weak scale is determined, 
through Renormalisation Group (RG) evolution, by specifying the following high scale parameters: $m_0$, 
the common scalar mass, $A_0$,  the common trilinear coupling and $M_{1/2}$, 
the universal gaugino mass. The remaining low energy parameters of the model are $\tan \beta$, the ratio of the Higgs vacuum expectation values (vev) and the sign of $\mu$. 
The superpotential of the CMSSM incorporating the seesaw mechanism~\cite{seesaw} can
be written as 

\begin{equation}
\label{eq1}
W = W_{Y_Q} + h^e_{ij} L_i e^c_j H_1 + h^\nu_{ij} L_i \nu^c_j H_2 + M_{R_{ij}} 
\nu_i^c \nu_j, 
\end{equation} 
where the leptonic part has been detailed, while the quark Yukawa
couplings and the $\mu$ parameter are contained in $W_{Y_Q}$. $i,j$ are
generation indices.
$M_{R}$ represents the (heavy) Majorana mass matrix for the right-handed
neutrinos.  Eq.(\ref{eq1}) leads to the standard seesaw 
formula for the (light) neutrino mass matrix 
\begin{equation}
{\mathcal M}_\nu = - h^\nu M_{R}^{-1} h^{\nu~T} v_2^2,
\end{equation}
where $v_2$ is the vacuum expectation value (VEV) of the up-type
Higgs field, $H_2$. Under suitable conditions on $h^\nu$ and $M_R$
the correct mass splittings and  mixing angles in $\mathcal{M}_\nu$ can be obtained. 
Detailed analyses deriving these conditions are already present in the 
literature \cite{seesawreviews}.  Additionally, the decoupled heavy right handed 
neutrinos also leave their imprints on the soft sector in terms of flavour 
violating entries \cite{fbam} which can be probed in LFV experiments. 

The amount of lepton flavour violation generated by the SUSY seesaw crucially
depends on the flavour structure of $h^\nu$ and $M_R$, the new sources of 
flavour violation which do not appear in the CMSSM. This dependence can be 
clearly seen in the Renormalisation Group-induced entries in 
the slepton mass matrices at the weak scale:
\begin{equation}
\label{rgemi}
(m_{\tilde L}^2)_{ij}\approx -{3 m_0^2+A_0^2 \over 8 \pi^2}
\sum_k (h^\nu_{ik} h^{\nu *}_{jk}) \ln{M_{GUT} \over M_{R_k} }\,,\qquad i\neq j\,,
\end{equation}
where $M_R$ is the scale of the right handed neutrinos. 
From above it is obvious that if either the neutrino Yukawa couplings
or the flavour mixings present in $h^\nu$ are very tiny, the strength
of LFV will be significantly reduced. Given that there are several possible
regions in the seesaw parameter space which may generate the observed mixing in
the neutrino sector, one has to resort to some assumptions on $h^\nu$
and on $M_R$ to make a quantitative analysis of LFV within this model.

In the present work, we take the view that the presence of non-zero neutrino
masses modifies the CMSSM in such a way that the operative seesaw mechanism
significantly maximises LFV leading, in the 
near future, to another sound discovery road for SUSY. To this extent, 
let us make the following assumptions \cite{assumptions,oscar,hisano}: \\
(a) At least one of the neutrino Yukawa couplings is of $\mathcal{O}(1)$. For definiteness,
we choose it to be the third eigenvalue, $h^\nu_3$, setting it equal to the top
quark Yukawa, $h_t$. Assuming neutrino masses are hierarchical, this automatically sets 
the largest eigenvalue of $M_R$ to be rather close to $M_{GUT}$, around $10^{14} \div 10^{15}$ 
GeV. The second largest and the smallest eigenvalues  $h^\nu_{2,1}$ can be left unspecified 
as far as the dominant contribution to lepton flavour violation is concerned. However, for
a complete neutrino mass model, one needs to specify these eigenvalues too.\\
(b) The matrix which diagonalises the product $h^\nu h^{\nu~ \dagger}$ has either
a CKM-like structure with small mixing or a MNS like structure with large mixing.   
Notice that the mixing present in $h^\nu h^{\nu ~\dagger}$ determines 
the amount of LFV, as given by Eq.(\ref{rgemi}). Together with assumption (a),
the case with small CKM-like mixing gives a ``worst case'' scenario, whereas the case with large MNS-like mixing provides a ``best case'' for LFV. 

As far as neutrino masses and mixings are concerned, assumptions (a) and (b) 
have been extensively studied in the literature, and it has been shown that they 
may lead to phenomenologically viable models for neutrino masses~\cite{oscar}. 
The relevant references can be found in the reviews listed in Ref.~\cite{seesawreviews}. 
Given these assumptions, the LFV mass-insertions at the weak scale 
in the CMSSMRN are given, for the worst and best case respectively as:

\noindent 
\textit{Worst case}:
\begin{eqnarray}
\label{wcmi1}
(m_{\tilde L}^2)_{21}&\approx& -{3 m_0^2+A_0^2 \over 8 \pi^2}~
h_t^2 V_{td} V_{ts} \ln{M_{GUT} \over M_{R_3}} + \mathcal{O}(h^\nu_2)^2, \\
\label{wcmi2}
(m_{\tilde L}^2)_{32}&\approx& -{3 m_0^2+A_0^2 \over 8 \pi^2}~
h_t^2 V_{tb} V_{ts} \ln{M_{GUT} \over M_{R_3}} + \mathcal{O}(h^\nu_2)^2, \\
\label{wcmi3}
(m_{\tilde L}^2)_{31}&\approx& -{3 m_0^2+A_0^2 \over 8 \pi^2}~
h_t^2 V_{tb} V_{td} \ln{M_{GUT} \over M_{R_3}} + \mathcal{O}(h^\nu_2)^2.
\end{eqnarray}
\noindent 
\textit{Best case}:
\begin{eqnarray}
\label{bcmi1}
(m_{\tilde L}^2)_{21}&\approx& -{3 m_0^2+A_0^2 \over 8 \pi^2}~
h_t^2 U_{e 3} U_{\mu 3} \ln{M_{GUT} \over M_{R_3}} + \mathcal{O}(h^\nu_2)^2, \\
\label{bcmi2}
(m_{\tilde L}^2)_{32}&\approx& -{3 m_0^2+A_0^2 \over 8 \pi^2}~
h_t^2 U_{\mu 3} U_{\tau 3} \ln{M_{GUT} \over M_{R_3}} + \mathcal{O}(h^\nu_2)^2, \\
\label{bcmi3}
(m_{\tilde L}^2)_{31}&\approx& -{3 m_0^2+A_0^2 \over 8 \pi^2}~
h_t^2 U_{e 3} U_{\tau 3} \ln{M_{GUT} \over M_{R_3}} + \mathcal{O}(h^\nu_2)^2, 
\end{eqnarray}
where $V$ is the CKM matrix and $U$ is the 
leptonic `PMNS' mixing matrix.
It is now obvious that the seesaw generated mass insertions above lead
to various LFV processes, like rare leptonic radiative decays,
$\mu \to e $ conversion in nuclei \cite{okada}, $\tau \to 3 \mu$~\cite{t3mu}. 
One of the features which stands out  is
that in the best case scenario, the amplitude for any LFV process involving $\mu \to e $ 
transitions {\em depends} on the neutrino mixing angle, $U_{e3}$, of which very little 
experimental information is known, except for an upper bound (See for ex: Ref.~\cite{ue3importance}). 
The same statement is also true for $\tau \to e$ transitions. The $\tau \to
\mu$ transitions are instead $U_{e3}$-independent probes of SUSY, whose importance was first
pointed out in Ref.~\cite{kingblazektmg}.
Otherwise, the leptonic flavour violating transitions are now completely determined only
in terms of the CMSSMRN parameters. 

To appreciate the phenomenological impact of these rare transitions, we list below
the present and upcoming experimental limits on the $l_j \to l_i, \gamma$ decays.\\

\noindent\textit{Present limits}:\\
\begin{center}
\begin{tabular}{rclc}
$BR(\mu \to e \gamma)$ &$\leq$& $1.2\times 10^{-11}$ & \cite{mega}\\
$BR(\tau \to \mu \gamma)$ &$\leq$& $3.1\times 10^{-7}$ & \cite{belletmg}\\
$BR(\tau \to e \gamma)$ &$\leq$& $3.7\times 10^{-7}$ & \cite{belletalk}
\end{tabular}
\end{center}

\noindent\textit{Upcoming limits}:\\
\begin{center}
\begin{tabular}{rclc}
$BR(\mu \to e \gamma)$ &$\leq$& $10^{-13} \div 10^{-14}$ & \cite{psi}\\
$BR(\tau \to \mu \gamma)$ &$\leq$& $10^{-8}$ & \cite{belletalk}\\
$BR(\tau \to e \gamma)$ &$\leq$& $10^{-8}$ & \cite{belletalk}
\end{tabular}
\end{center}

In the next Sections, we will apply these bounds to constrain the CMSSMRN parameter space.\\ 

\subsection{The Three Roads }\label{sec:darkmatter}

The seesaw mechanism operative within our model does not modify the SUSY mass 
spectrum of the CMSSM significantly. Thus except for LFV and other rare processes, 
the phenomenology of the CMSSM is very similar to that of the CMSSMRN. Let us briefly summarise the status 
of the mentioned three roads and spell out the various constraints we consider in the present work.\\

(i) {\bf Collider Searches:}~  Direct collider searches will give conclusive 
evidences for the existence of low energy supersymmetry. The searches conducted 
at LEP already provide strong bounds on the CMSSM parameter space. 
The main constraints are \cite{lepconstraints}:\\ 

\noindent (a) The SM-like light Higgs boson mass:

\begin{equation}
\label{lephiggsbound}
m_h ~\geq~ 114.1 ~{\rm GeV}, \ {\rm and}
\end{equation} 
(b) The lightest chargino mass:
\begin{equation}
\label{lepcharginobound}
 m_{\chi_1^\pm} \geq 103~ {\rm GeV}.
\end{equation}   
These bounds significantly constrain the CMSSM parameter space, especially
in the low $\tan \beta$ and low mass spectrum (``{\em bulk} '') region. Soon, 
more extended CMSSM regions will also be probed by the Fermilab Tevatron\footnote{It has been 
recently shown that parameter space up to $m_0 \leq 200$ GeV and $M_{1/2} \leq 250 $ GeV could 
be probed for a low tan $\beta$ value assuming that the integrated luminosity of the 
Tevatron reaches 25 $fb^{-1}$  \cite{baertevatron}.}. 
A by far larger portion of the CMSSM parameter space will be probed at the CERN Large
Hadron Collider (LHC)\footnote{Here, in terms of $m_0$ and $M_{1/2}$, 
the maximal region covered by LHC, at an integrated luminosity of $\sim100\ {\rm fb}^{-1}$, will be up to 1400 GeV in $M_{1/2}$ for $m_0 \leq $ 1 TeV.  
At larger $m_0 \gtrsim 5 $ TeV, the maximal reach in $M_{1/2}$ will instead be around 700 GeV \cite{Baer:2003wx}.}.\\

(ii) {\bf Rare processes}:
Low-energy supersymmetry can also manifest itself in various flavour and CP
violating processes (see \cite{gabbiani} for a review). In the CMSSM, where
for most of the processes there is no significant deviation from the SM predictions, 
the main constraint on the parameter space stems from $BR(b\to s, \gamma)$ \cite{minimalfv} (see also \cite{borzumati} for recent progresses). 
At present, combining the theoretical and experimental uncertainties, this 
constraint is given by \cite{bsgexp}:
\begin{equation}\label{eq:bsg}
2.16<BR(b\rightarrow s\gamma)\times 10^4<4.34
\end{equation}
The other main constraints come from the SUSY contributions to the anomalous magnetic 
moment of the muon, $\delta a_\mu$, and from the Higgs mediated $B_s \to \mu^+ \mu^-$. 
It is known that the anomalous magnetic moment of the muon is capable of putting 
severe constraints on the supersymmetric parameter space. However, at present there 
are large theoretical and experimental uncertainties involved in evaluating the SM 
contribution to $\delta a_\mu$. The other significant constraint might come from the upper limit 
on the $BR(B_s \to \mu^+ \mu^-)$ from Tevatron \cite{Hmedmu1,Hmedmu2}, but again as far
as the CMSSM is concerned, $BR(b\to s, \gamma)$ is mostly a much stronger bound. We postpone
the consideration of these effects to a future work.\\

(iii) {\bf Neutralino Relic Density and Dark Matter Searches}: In the CMSSM the lightest neutralino $\tilde\chi_1$ is the lightest supersymmetric 
particle (LSP) over a very wide range of parameters, providing an ideal particle 
candidate for cold dark matter. $\tilde\chi_1$ is practically always, to a 
high degree of purity, a {\em bino}, except for a thin strip close to the 
parameter space area where radiative electroweak supersymmetry breaking conditions are
 not fulfilled, named {\em focus point region} \cite{focuspoint}. 

The current ``state-of-the-art'' of supersymmetric dark matter studies 
faces on the one side an impressive progress in observational cosmology, 
which recently led to an unprecedented accurate determination of the amount of cold dark matter in the universe \cite{wmap}:
\begin{equation}\label{eq:cdmabundance}
\Omega_{\rm CDM}h^2=0.1126^{+0.008}_{-0.009},
\end{equation}
where $\Omega_{\rm CDM}$ represents the cold matter density of the universe
and $h$ parameterises the present Hubble rate, $H_0 = 100 h~\mbox{Km}~\mbox{sec}^{-1} 
\mbox{Mpc}^{-1}$, with $h~=~0.72 \pm 0.08$. On the 
other hand, extremely accurate numerical packages allow to compute, with 
a typical accuracy of 1\% or better, the neutralino thermal relic 
abundance for a given supersymmetric
setup \cite{Gondolo:2002tz,Edsjo:2003us,Belanger:2001fz}.\\

We will require that neutralinos either totally or partly contribute to the
 inferred dark matter density. We point out that in most models of low energy 
SUSY it turns out that the computed amount of thermal relic neutralinos is typically 
larger than what indicated in Eq.~(\ref{eq:cdmabundance}) 
above\footnote{It should be noted that in several examples, the presence of 
{\em extra components} in the Universe energy density at neutralino decoupling, 
for instance a scalar field in a Brans-Dicke-Jordan 
cosmology \cite{Kamionkowski:1990ni}, a 
quintessential field \cite{Salati:2002md,Rosati:2003yw,Profumo:2003hq}, or the shear 
energy density associated to primordial anisotropies \cite{Kamionkowski:1990ni}, 
only strongly {\em enhances} the predicted neutralino abundance: for instance, the
 presence of a quintessential scalar field may lead to enhancements in the
 neutralino relic density up to {\em six orders of magnitude} 
 \cite{Profumo:2003hq}.}. With this in mind, and allowing for the presence of extra DM components beside neutralinos, we will only consider as a constraint the 95\% C.L. upper bound on the thermal relic abundance of 
neutralinos derived from Eq.(\ref{eq:cdmabundance}):
\begin{equation}
\label{neutcdm}
\Omega_{\tilde\chi_1}h^2<0.129.
\end{equation}

As regards prospects for direct and indirect detection of
neutralino dark matter within the constrained MSSM,
\cite{ellisdirect,otherdirect,Bertin:2002sq}
 it has been shown that the most favorable parameter space points lie
in the focus point-hyperbolic branch region.
For instance, regarding the muon flux from neutralino pair-annihilation in the center of
the Earth or of the Sun, one of the most promising
indirect detection strategies, planned neutrino telescopes
\cite{Bertin:2002sq,antares,icecube} will probe, within the CMSSM,
practically only regions at low neutralino masses and large
higgsino content. On the other hand, it has been shown
\cite{ellisdirect} that direct detection
spin-dependent searches have typical projected sensitivities lying
orders of magnitude far from the expected signal. As regards,
finally, direct detection through spin-independent interactions,
once again promising signals are predicted only in the low
neutralino masses and/or large higgsino fractions parameter space
points of the CMSSM.
To summarise, at present and future facilities, direct and
indirect dark matter searches will compete with the LHC only in a
very limited CMSSM parameter space region, in the focus point
strip at moderate values of $M_{1/2}$ , and therefore at low
neutralino masses.\\
\\

Combining all the constraints given by 
Eqs.(\ref{lephiggsbound}, \ref{lepcharginobound}, \ref{eq:bsg}) and in particular  Eq.~(\ref{neutcdm}),
apart from the well-known {\em bulk region}, at very low soft breaking masses, which will be 
thoroughly probed by Tevatron and the LHC, three main regions survive in the CMSSM parameter space: 
\begin{itemize}
\item \textit{The Stau Coannihilation Regions:} If the lightest stau and the lightest neutralino 
are quasi degenerate in mass, efficient stau-stau as well as stau-neutralino 
(co-)an\-ni\-hi\-la\-tions~\cite{griest} significantly contribute  in suppressing the thermal 
$\chi_1$ relic abundance\footnote{Similar mechanisms with the lightest stop are possible in 
regions with very large trilinear scalar couplings~\cite{Edsjo:2003us,Boehm:1999bj}. We do not consider 
these regions in the present work.}. Most of these regions will be accessible to the LHC, at any value of $\tan\beta$, with the possible exception of very large values of $M_{1/2}$. 
\item \textit{Funnel Regions:} In these regions, the bino-bino annihilation cross section 
is greatly enhanced through resonant $s$-channel exchange of the heavy neutral Higgses $A$ and $H$. 
The conditions required for the resonant enhancement are a large value for $\tan\beta $ and $\mu<0$, 
which are necessary to fulfill the relation $ 2 m_{\chi_1^0} \approx m_A$. Here LHC might cover the parameter 
space corresponding, roughly, to $M_{1/2} \leq 1$ TeV.
\item \textit{Focus point or Hyperbolic Branch Regions:} These regions are narrow zones corresponding
to large values of $m_0$, yielding a {\em low} value for the $\mu$ parameter. The focus point region 
lies close to where radiative electroweak symmetry breaking (EWSB) is not valid, and is also relatively 
unstable numerically. The main feature is that a non-negligible higgsino fraction in the lightest 
neutralino is produced. The LHC reach of this region is pretty limited, owing to the large values 
of $m_0$ and $M_{1/2}$, yielding very heavy gluinos and squarks. 
\end{itemize}

The `smallness' of the surviving regions in the CMSSM is a result of the strict 
universality assumptions on the soft masses at the high scale within the CMSSM. 
It has been shown that, relaxing some assumptions about the Higgs 
\cite{Ellis:2002wv,Ellis:2002iu} or the sfermion \cite{spnonuniv} soft 
SUSY breaking mass universality, other coannihilation partners may arise, 
such as the sneutrino \cite{Ellis:2002wv,spnonuniv} or the bottom
squark \cite{spnonuniv}, leading to larger allowed regions in the parameter space
compared to the present situation.

We compute the neutralino relic density through the {\tt darkSUSY} package \cite{Gondolo:2002tz}, 
which merges the accurate numerical SUSY RGE's evolution package {\tt ISASUGRA} \cite{Baer:2003mg} 
with the state-of-the-art technique to treat the thermal averaging of cross sections, coannihilations, 
resonances and threshold effects \cite{Edsjo:2003us}. To evaluate the lepton flavour violating rates we use the complete formul\ae of Ref.~\cite{hisano}, in the mass-eigenstate
basis.

\section{The Canonical ($m_0~-~ M_{1/2}$) Plane and LFV }\label{sec:paramspace}

We present our results for the effects of LFV on the CMSSMRN parameter space 
in the $(m_0 - M_{1/2})$ plane, as customary in CMSSM studies. We then
concentrate on the regions of the parameter space which satisfy all the cosmo-phenomenological constraints outlined in the previous Section, which are plotted as green contours.
As a starting point, in this Section we present the results for the `best case'
scenario in the plane $(M_{1/2},m_0)$ for a fixed value of $\tan\beta$ and $A_0=0$, and 
a given sign of $\mu$. We will always show in what follows both a {\em small} and a {\em large} $\tan\beta$ case, respectively setting $\tan\beta=10$ and 50 (for thourough discussions about the $\tan\beta$ dependence in LFV processes see Ref.~\cite{lfv2,hisano}), while we devote sec.~\ref{sec:A0} to the question of a non-vanishing trilinear coupling $A_0$.

\begin{figure*}
\begin{center}
\includegraphics[scale=0.7]{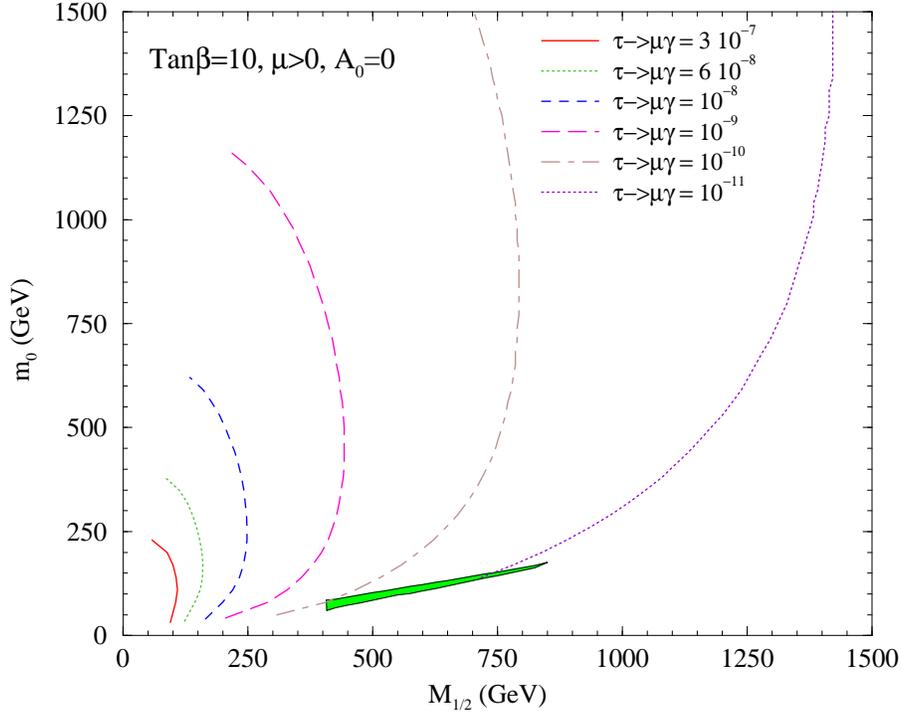}
\end{center}
\caption{\em The area shaded in green shows the parameter space, in the 
$(M_{1/2},m_0)$ plane, allowed by all phenomenological and cosmological 
constraints, for $\tan\beta=10$, $\mu>0$ and $A_0=0$. The lines correspond 
to various isolevel curves at $\brtm=$$3\cdot 10^{-7}$, $6\cdot 10^{-8}$,
 $10^{-8}$, $10^{-9}$, $10^{-10}$, $10^{-11}$. The starting point of each 
line is dictated by the condition $m_{\tilde\chi_1}=m_{\tilde\tau}$, which 
fixes the lowest possible value of $m_0$. The isolevel 
curves terminate where radiative electroweak SUSY breaking (EWSB) conditions 
cannot be fulfilled, therefore giving the largest possible value of 
$m_0$.}\label{fig:PS10}
\end{figure*}

In Fig.~{\ref{fig:PS10}} we show the isolevel curves of constant $\brtm$ for $\tan\beta=10$ and $\mu>0$.
From the figure, we see that the  present experimental limit already starts
probing the region of the parameter space at low $m_0$ and $M_{1/2}$, which is likely to 
be tested at Tevatron. The green patch representing 
the coannihilation region intersects the isolevels corresponding to $\brtm$ between $10^{-10}$ 
and $10^{-11}$, a level of sensitivity likely to be achieved only at dedicated $\tau$-factories.
In this low tan $\beta$ coannihilation region, which will be entirely probed at the LHC, LFV does not significantly impact on the quest for SUSY.  

\begin{figure*}
\begin{center}
\includegraphics[scale=0.7]{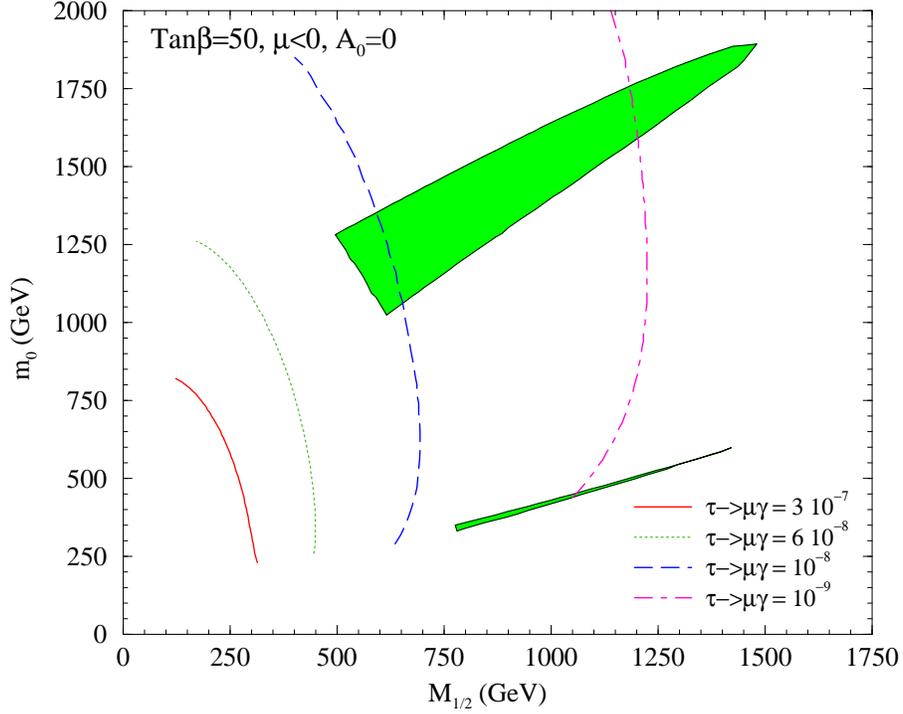}
\end{center}
\caption{\em The area shaded in green shows the parameter space, in the 
$(M_{1/2},m_0)$ plane, allowed by all phenomenological and cosmological 
constraints, for $\tan\beta=50$, $\mu<0$ and $A_0=0$. The upper area at 
large $m_0$ represents the funnel region. The lines correspond to various isolevel curves 
at $\brtm=$ $3\cdot 10^{-7}$, $6\cdot 10^{-8}$, $10^{-8}$, $10^{-9}$. 
As in Fig.~\ref{fig:PS10}, each line starts at  
$m_{\tilde\chi_1}=m_{\tilde\tau}$ and ends where REWSB conditions are no 
longer fulfilled.}\label{fig:PS50}
\end{figure*}

The situation changes however for the large  $\tan\beta$ region, where one expects a large enhancement in the decay rates (which roughly scale as $\tan^2\beta$). 
In Fig.{\ref{fig:PS50}}, we show $\brtm$ iso-contours for $\tan\beta=50$. Notice that we have now
chosen $\mu<0$ so that the A-pole funnel regions (see figure caption) also appear within the allowed parameter space. Here we see that even the present experimental bound on $\brtm$ 
can exclude values of $m_0$ up to $800\,$GeV and up to $300\,$GeV in $M_{1/2}$. 
The intersection of the isolevel curves with the  coannihilation region and with the funnel takes 
place for $\brtm\approx 10^{-8}$ and extends beyond $10^{-9}$, well within the reach of 
present as well as proposed B-factories.

\begin{figure*}\label{PSME50}
\begin{center}
\includegraphics[scale=0.7]{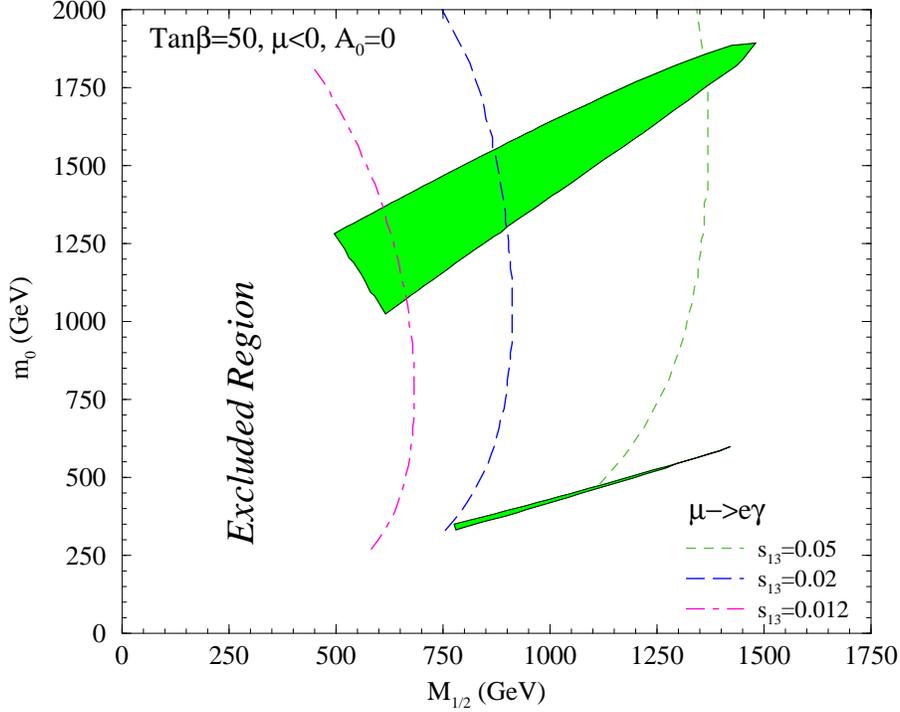}
\end{center}
\caption{\em The area shaded in green shows, as in Fig.~\ref{fig:PS50}, 
the allowed parameter space, in the $(M_{1/2},m_0)$ plane, for $\tan\beta=50$, 
$\mu<0$ and $A_0=0$. The lines correspond this time to the {\em exclusion curves} 
dictated by the current experimental bound on $\brmu$, for various values of 
$s_{13}=0.05,\ 0.02$ and $0.012$: points lying to the left of these curves are 
henceforth {\em ruled out} by the current bound $\brmu<1.2\cdot 10^{-11}$. The 
extreme case $s_{13}=0.2$ would rule out the entire parameter space, and is not 
shown. Again, as in the preceding figures, each line begins at  
$m_{\tilde\chi_1}=m_{\tilde\tau}$ and ends where EWSB conditions are no longer 
fulfilled.}\label{fig:PSME50}
\end{figure*}

\begin{figure*}
\begin{center}
\includegraphics[scale=0.7]{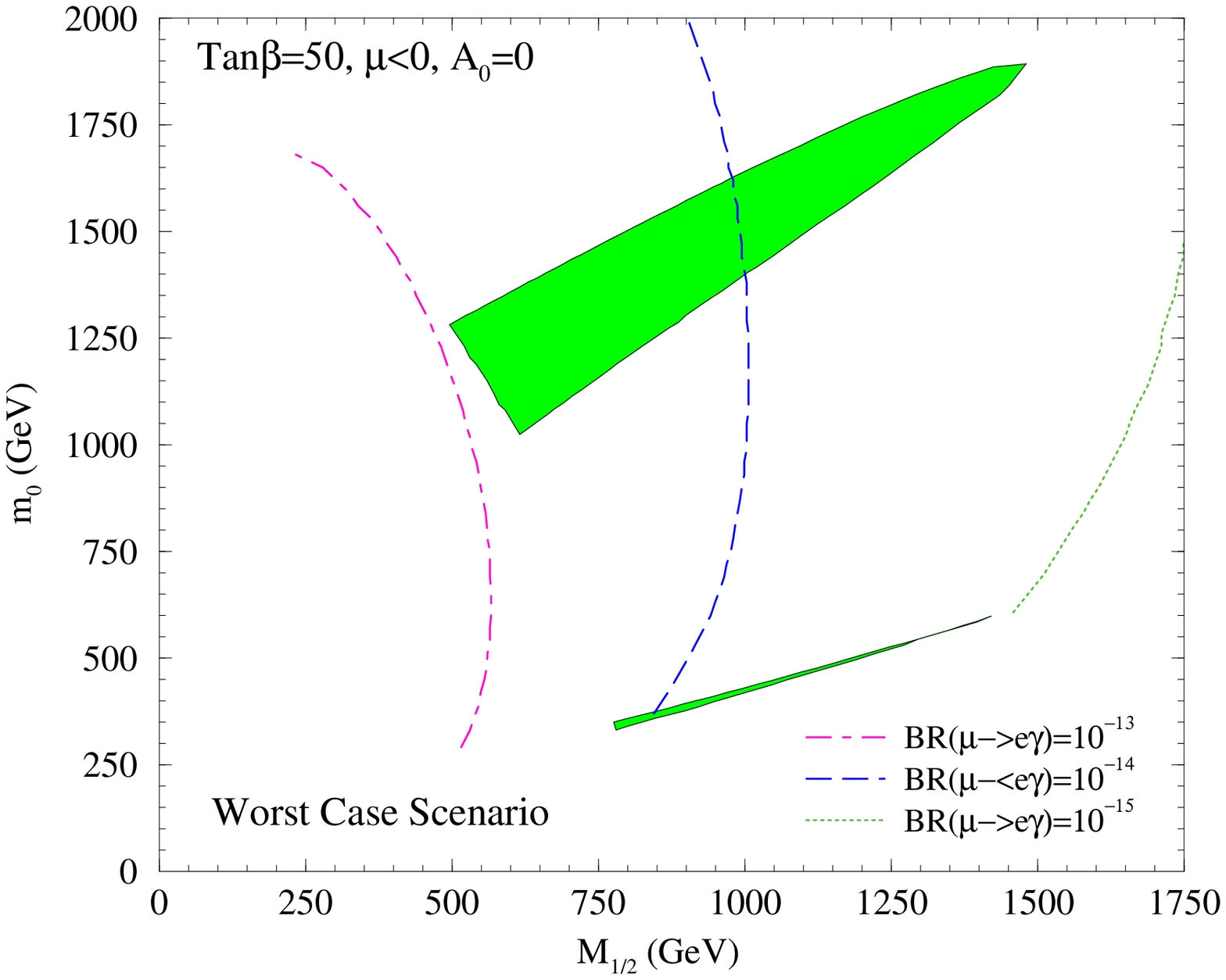}
\end{center}
\caption{\em Different isolevel curves of $\brmu$ for the {\em worst case scenario}, \ie\ corresponding to a CKM-like mixing in the neutrino mixing matrix, on the same parameter space as in Fig.~\ref{PSME50}, at  $\tan\beta=50$, 
$\mu<0$ and $A_0=0$.}\label{fig:PS50WORST}
\end{figure*}

The amplitude of  $\mu\rightarrow e \gamma$ is $U_{e3}$ dependent, as mentioned earlier. In Fig.~{\ref{fig:PSME50}}, we plot isolevels corresponding 
to the present experimental limit on $\brmu$ for various values of $U_{e3}$. Notice 
that we use here as a parameter the mixing angle $s_{13}=e^{-i\delta}U_{e3}$. 
We choose again $\tan\beta=50$ and $\mu<0$. The results are remarkable. 
Taking $s_{13}$=0.2, close to the present experimental upper bound from CHOOZ, 
would rule out the whole parameter space. This fact can be interpreted either as a negative result 
for a large $U_{e3}$ or for a large value of $\tan\beta$. In any case, it implies relevant
 consequences for supersymmetric seesaw models. In the plot, we show values starting from $s_{13} = 0.05$.
We see that a value of $s_{13}$ ten times smaller than the present bound ($s_{13}=0.02$) can 
still exclude $M_{1/2}<750\,$GeV for any value of $m_0$. 

Finally, we show in Fig.~\ref{fig:PS50WORST} the isolevel curves corresponding to various values of $\brmu$ in the {\em worst case scenario}, which, we recall, features a CKM-like mixing in the neutrino Yukawa couplings. Though $\tan\beta$ is fixed to a large value, 50, LFV rates are nonetheless rather suppressed: the current experimental bound, for instance, does not give any constraint on the SUSY  parameter space. On the other hand, even in this case an improvement in the experimental sensitivity on $\brmu$ may lead to access a large part of the cosmo-phenomenologically viable regions. From this point on we will however only consider the {\em best case scenario}, because it typically gives more interesting constraints and elucidates the possible relevance of $U_{e3}$ for LFV.

\begin{figure*}
\begin{center}
\includegraphics[scale=0.6 ]{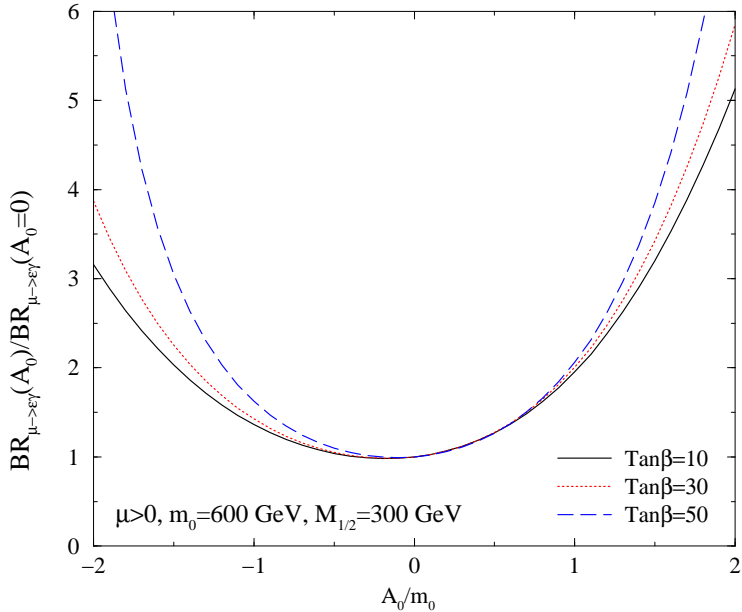}
\end{center}
\caption{\em The dependence of $\brmu$ on the trilinear coupling $A_0$, for the 
particular choice $\mu>0$, $m_0=600\ {\rm GeV}$ and $M_{1/2}=300 {\rm \ GeV}$. 
The three lines correspond to the ratio of $\brmu$ computed at a given 
$A_0\neq0$ and its value at $A_0=0$, for three different values of 
$\tan\beta=10,\ 30,\ 50$.}\label{fig:A0}
\end{figure*}
\subsection{The Dependence on $A_0$}\label{sec:A0}
Before proceeding to the next section, we pause to comment on the implications of 
a non-zero trilinear coupling on the LFV results.  As mentioned above, so far 
the parameter $A_0$ has been set to zero. We argue here that any 
value $A_0\neq 0$ would give a larger branching ratio and therefore the case $A_0=0$ 
provides a {\em lower bound}. 

A non-zero $A_0$ affects the 
amplitudes of $\l_j \to l_i \gamma$ decays in three ways: (i) The first effect comes from RG scaling. 
From Eq.(\ref{rgemi}), we see that the RG-generated off-diagonal flavour violating mass-insertions 
are proportional to  $A_0^2$. This is the dominant effect coming from a non-zero $A_0$, and it is 
independent on its sign. (ii) Another way in which $A_0$ affects the branchings is 
through modifications to the particle spectrum. In particular, the parameter $\mu$ is sensitive to 
the value and to the sign of $A_0$. However, this produces only a small effect in the branchings ratios, 
because $\mu$ in the CMSSM is associated with the heavier chargino and the heaviest neutralino
 states. This effect is responsible of the manifest, though small, asymmetry for positive and negative values of $A_0$ (see Fig.~\ref{fig:A0}). (iii) Last, the left-right mixing elements of the slepton mass matrix contain $A_0$ in 
addition to the $\mu \tan \beta$ factor. This `left-right' mixing may play an important role 
in enhancing the decay amplitudes \cite{hisano,largetanbeta}. However, it turns out that for most 
of the parameter space, the main contribution to this mixing is from the  $\mu \tan \beta$ terms rather 
than from non-zero values of $A_0$.\\
As a quantitative reference, in Fig \ref{fig:A0} we show, as a 
function of $A_0/m_0$, the ratio between $\brmu$ and its value at $A_0=0$, for 
different values of $\tan\beta$. In agreement with what stated above, the 
minimum value of the branching is obtained for $A_0=0$, and there is only a 
slight dependence with the sign of $A_0$. Thus all the predictions  presented
in this and the later section have actually to be regarded as \textit{lower bounds}
on LFV rates. Let us finally point out that the overall variation induced by a non-zero value of 
the $A_0$ parameter lies within less than one order of magnitude. Finally, we remark that we numerically verified that these conclusions do not depend qualitatively on the particular parameter space point under consideration.

\section{Coannihilations, Funnels and Focus Point}\label{sec:predictions}

LFV processes constitute a meaningful constraint on SUSY models, as the results 
for the `best case'  scenario presented in the preceding section manifestly show. 
We will demonstrate that sometimes they even {\em do better} than direct accelerator SUSY searches.
Given this situation, we now come back
to the question of the complementarity between the various roads leading to SUSY discovery, 
directly comparing the LHC reach with the prospects for LFV experiments. 
We therefore now proceed to detail on the impact of LFV rates for each of the three 
allowed regions of the CMSSMRN, for benchmark $\tan\beta$ values. Namely, we will concentrate 
on (a) the coannihilation region, (b) the funnel region and finally (c) the focus point regions, 
as mentioned at the end of Sec.~3. 

\subsection{Coannihilation Strips}\label{sec:coannihilation}

The defining condition for this region is $m_{\tilde\chi_1}\simeq m_{\tilde\tau_1}$,
where $\tilde{\tau_1}$ denotes the lighter stau. 
In the present section, we stick to the points in the $(m_0 - M_{1/2})$ plane which saturate 
the limiting condition  $m_{\tilde\chi_1}= m_{\tilde\tau_1}$, maximising the extension of the coannihilation strip. We further
choose two values for $\tan\beta$, 10 and 50, setting, in the first case 
$\mu>0$, while in the second $\mu<0$. The scalar trilinear coupling $A_0$ is 
always set to zero (see Sec.\ref{sec:A0}).
\begin{figure*}
\begin{center}
\includegraphics[scale=0.7]{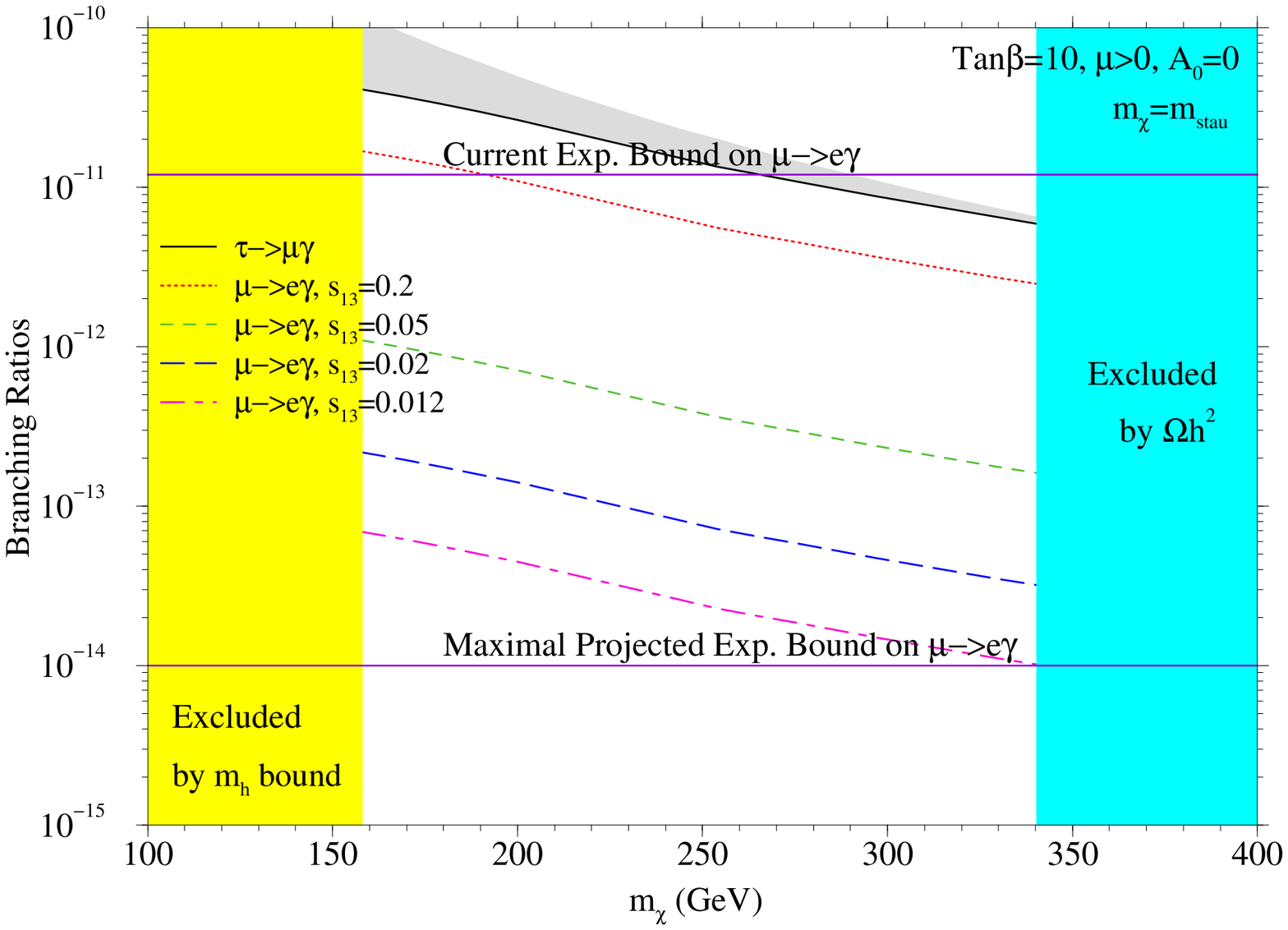}
\end{center}
\caption{\em $\brtm$ and $\brmu$, for various values of 
$s_{13}=0.2,\ 0.05,\ 0.02$ and $0.012$, along the line, in the 
$(M_{1/2},m_0)$ plane for $\tan\beta=10$, $\mu>0$ and $A_0=0$, where 
$m_{\tilde\chi_1}=m_{\tilde\tau_1}$, \ie\ in the lowest part of the 
coannihilation strip. The yellow shaded area at low neutralino masses is 
ruled out by the LEP bound on $m_h$,  while points within the cyan shaded region 
at large neutralino masses give an $\Omega_{\tilde\chi_1} h^2$ value which exceeds
 the current WMAP constraint on the cold dark matter density. The region shaded 
in gray indicates the possible values for  $\brtm$ which one can obtain varying
 $m_0$ within the parameter space showed in Fig.~\ref{fig:PS10}. The shape of 
these shaded regions is analogous for the other lines referring to $\brmu$'s.
 We also show the current and projected sensitivities for $\brmu$. We stress 
that all the points showed in this plot are within the expected CERN LHC reach 
at an integrated luminosity $\sim\ 100\ {\rm fb^{-1}}$.}\label{fig:CO10}
\end{figure*}

In Fig.~\ref{fig:CO10} we show our predictions for $\brtm$ and for $\brmu$ in 
the $\tan\beta=10$, $\mu>0$ case. The yellow region dictates the lower bound on
the neutralino mass, provided by the LEP constraint on the mass of the lightest 
$CP$-even Higgs boson $m_h$ \cite{lepconstraints}. On the other hand, the cyan region 
gives the upper bound, dictated by the point
 where $m_{\tilde\chi_1}=m_{\tilde\tau_1}$ and $\Omega_{\tilde\chi_1}h^2=0.129$, 
\ie\ the maximal neutralino mass in the coannihilation strip compatible with dark 
matter constraints \cite{wmap}.
We also show the current and projected experimental sensitivity for $\brmu$. 
We stress that all the parameter space points shown in the plot will be 
{\em within the expected sensitivity of CERN LHC}, as the latter extends, for this 
value of $\tan\beta$, up to approximately $m_{\tilde\chi_1}\simeq550\ {\rm GeV}$ in the
coannihilation strip.
Clearly, for such a low value of $\tan\beta$, LFV rates are rather suppressed, 
and at present one can just exclude a narrow region at low neutralino masses 
provided $s_{13}$ is close to its present upper bound. Moreover, $\brtm$ lies 
at least two orders of magnitudes below the planned experimental 
sensitivity. Interestingly enough, in case the 
experimental sensitivity on $\brmu$ is lowered down to $10^{-14}$, it will be possible 
to detect, within this scenario, $\mu \to e \gamma$ for $s_{13}$ as low as $10^{-2}$.

The gray shaded band on the $\brtm$ line
is obtained by varying the parameters within the coannihilation region. Although the 
range of $m_0$ at a given 
$M_{1/2}$ (and therefore neutralino mass) is exceedingly tiny, the shaded area is somewhat large, the reason being that 
the iso-level curves of LFV rates, as shown in Fig.~\ref{fig:PS10}, are 
approximately parallel to the coannihilation strip.
\begin{figure*}
\begin{center}
($a$)\hspace{1cm}\includegraphics[scale=0.7]{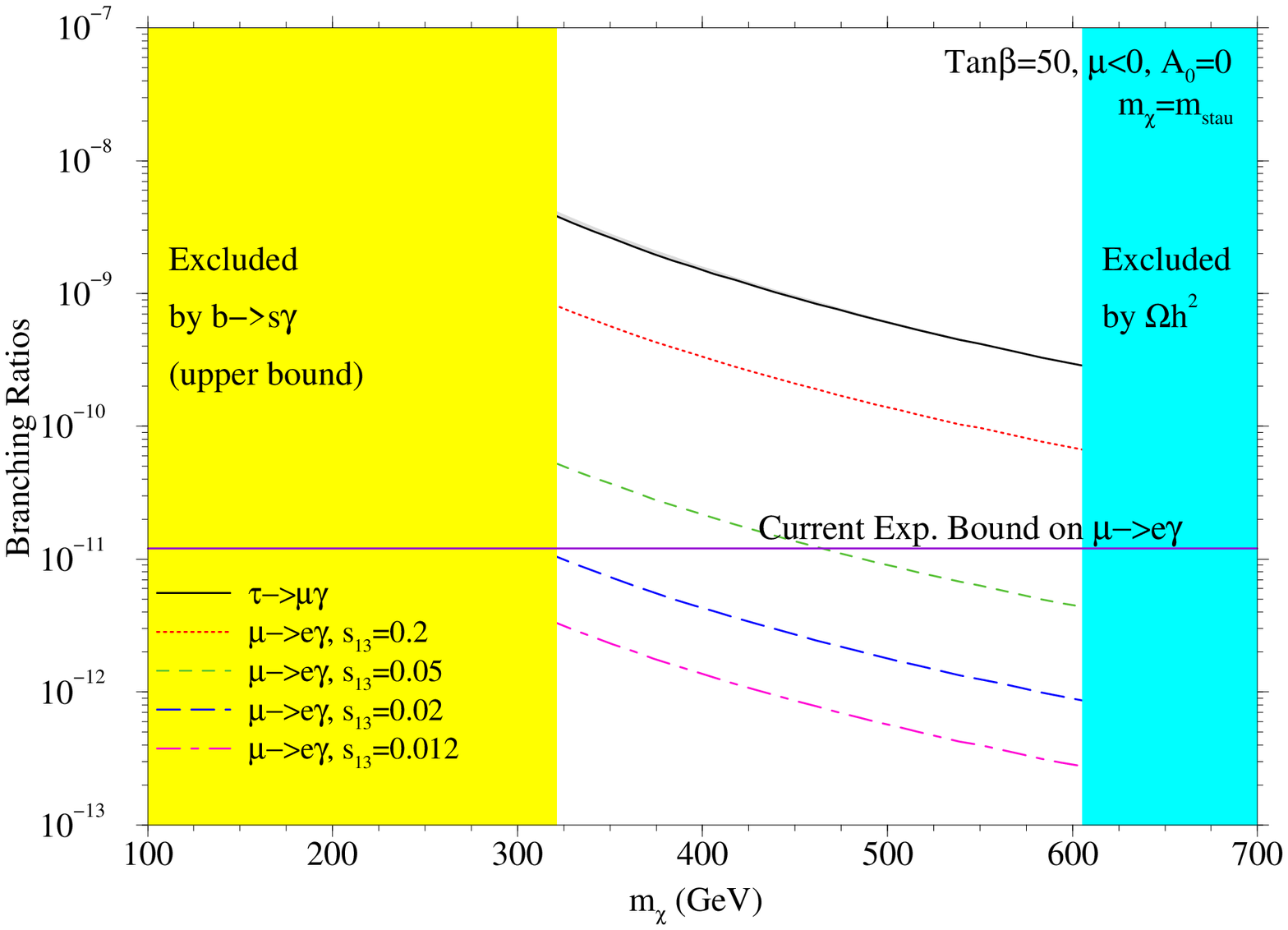}\\
($b$)\hspace{1cm}\includegraphics[scale=0.7]{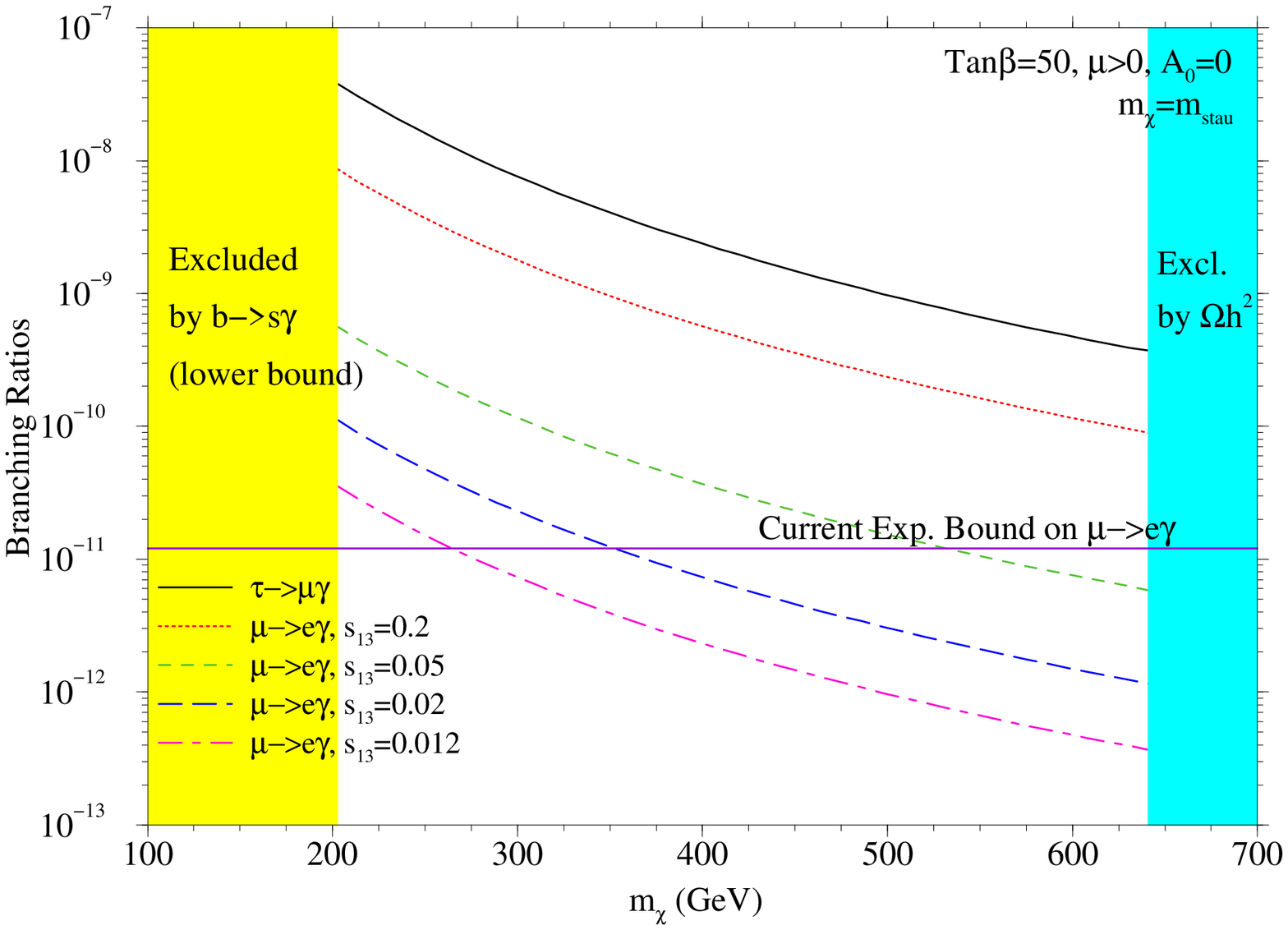}
\end{center}
\caption{\em $(a)$: $\brtm$ and $\brmu$, for various values of 
$s_{13}=0.2,\ 0.05,\ 0.02$ and $0.012$, along the line, in the $(M_{1/2},m_0)$ 
plane for $\tan\beta=50$, $\mu<0$ and $A_0=0$, where $m_{\tilde\chi_1}=m_{\tilde\tau_1}$, 
\ie\ in the lowest part of the coannihilation strip. 
The yellow shaded area at low neutralino masses is ruled out here by the 
inclusive $BR(b\rightarrow s\gamma)$ bound,  while points within the cyan 
shaded region at large neutralino masses give $\Omega_{\tilde\chi_1} h^2$ 
exceeding the current WMAP upper bound on CDM abundance. The current 
experimental bound on $\brmu$ is shown by an horizontal solid violet line. 
The region shaded in gray indicates the possible values for  $\brtm$ which one 
can obtain varying $m_0$ within the coannihilation parameter space showed in 
the lower green strip of Fig.~\ref{fig:PS50}. Also here, all the points should 
be within the expected CERN LHC reach at an integrated luminosity 
$\sim\ 100\ {\rm fb^{-1}}$. $(b)$: The same as in $(a)$, but with positive
 $\mu$.}\label{fig:CO50}
\end{figure*}

As regards the large $\tan\beta$ region, we pick the benchmark value $\tan\beta=50$, and choose $\mu<0$  in 
Fig.~\ref{fig:CO50} ($a$), again along the 
coannihilation strip. This time, since the isolevel curves intersect the 
coannihilation area almost orthogonally, the overall dependence 
on the $m_0$ spread is completely negligible, and the gray shaded area is vanishingly 
small (but depicted over the $\tau\rightarrow\mu\gamma$ line).
The lower bound on the neutralino mass is dictated by the upper bound on the 
inclusive BR($b\rightarrow s\gamma$), which for large $\tan\beta$ and negative 
sign of $\mu$ strongly limits the low mass region of the parameter space. On 
the other hand, due to stronger couplings in the relevant (co-)annihilation 
cross sections, the coannihilation strip extends up to rather large neutralino
 masses.
In the present case, the reach of LHC should approximately coincide with the 
upper bound on the neutralino mass shown in the plot \cite{Baer:2003wx}.
 Noticeably, for such a large value of $\tan\beta$, the current experimental 
upper bound on $\brmu$ happens to put severe constraints on $s_{13}$: 
we can for instance qualitatively conclude that if $\tan\beta$ is so large, 
then $s_{13}$ must be of the order $10^{-2}$ or less. The $s_{13}=0.2$ 
line turns out, for instance, to be {\em completely excluded} by the present 
experimental bounds.
Concerning future improvements on the experimental sensitivity on $\brmu$, 
we notice that all the lines showed in Fig.~\ref{fig:CO50} ($a$) will 
certainly be within future reach. On the contrary, the situation for $\brtm$ 
is not equally favorable, not even in this large $\tan\beta$ 
scenario. Yet, $\brtm$ of $\mathcal{O}(10^{-8})$ would start probing this
region.

Lastly, in Fig.~\ref{fig:CO50} ($b$) we show what would happen switching the sign 
of $\mu$ to positive values: the lower limit on $m_{\tilde\chi_1}$ is now 
given by the {\em lower} bound on BR($b\rightarrow s\gamma$), therefore 
excluding a smaller region, and the coannihilation strip 
is also slightly enlarged towards larger masses. The predictions for LFV rates are nevertheless not much
 affected, except for the parameter space which is overall much wider, thus 
leaving an appealing window, at low masses, where LFV processes are particularly 
large, and even $\brtm$ may lie within planned experimental sensitivities.
\begin{figure*}
\begin{center}
\includegraphics[scale=0.7]{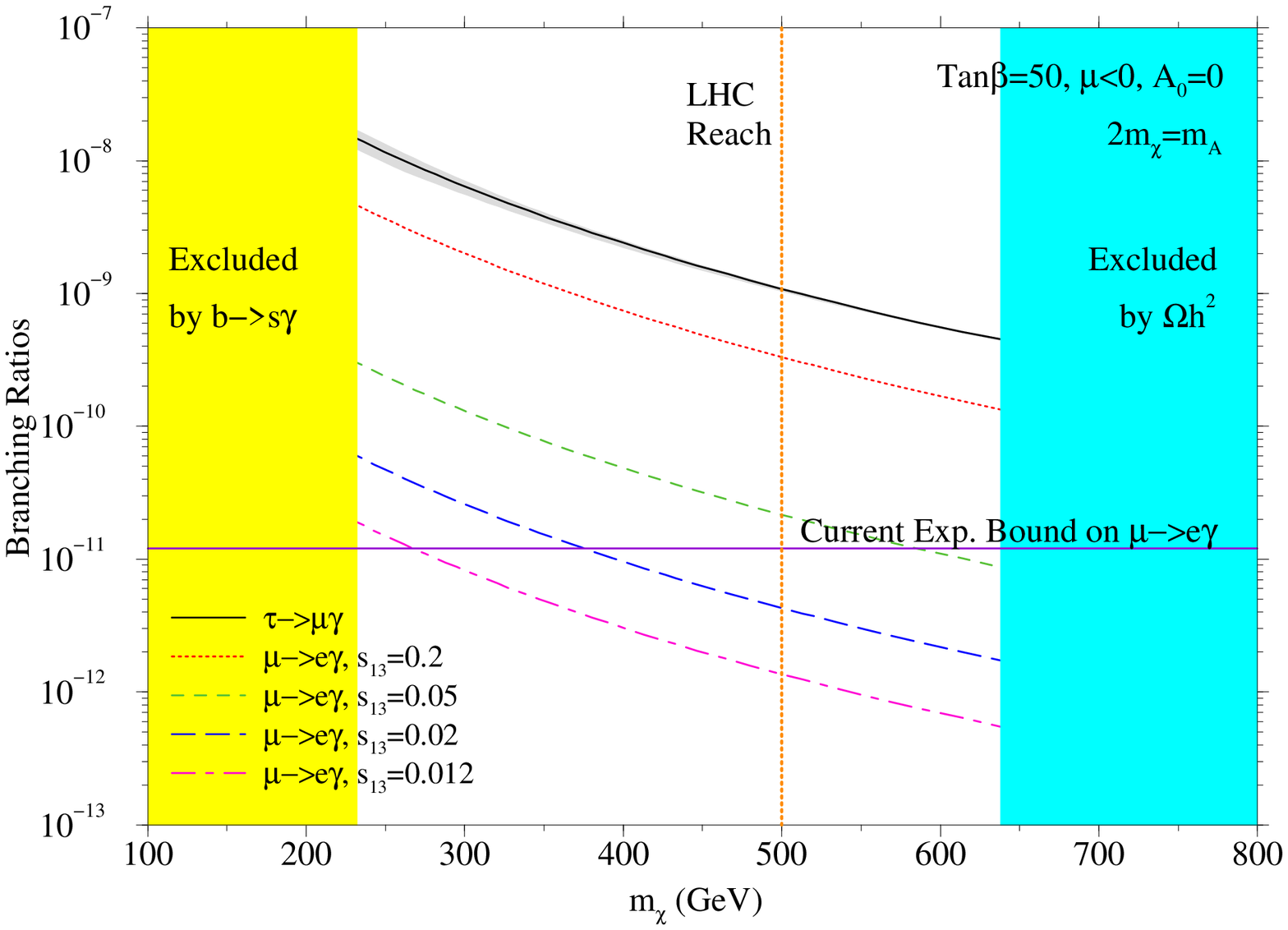}
\end{center}
\caption{\em $\brtm$ and $\brmu$, for various values of 
$s_{13}=0.2,\ 0.05,\ 0.02$ and $0.012$, along the line, in the 
$(M_{1/2},m_0)$ plane for $\tan\beta=50$, $\mu<0$ and $A_0=0$, 
where $2\cdot m_{\tilde\chi_1}=m_{A}$, \ie\ in the central part of the 
funnel region. The yellow shaded area at low neutralino masses is ruled out 
by the inclusive $BR(b\rightarrow s\gamma)$ bound,  while points within the 
cyan shaded region at large neutralino masses give an $\Omega_{\tilde\chi_1} h^2$ 
exceeding the current WMAP constraint on neutralino relic density. The region 
shaded in gray indicates the possible values for  $\brtm$ which one can obtain 
varying $m_0$ within the funnel parameter space as showed in the upper 
large green strip of Fig.~\ref{fig:PS50}. We also show the current and projected 
sensitivities to $\brmu$. The expected CERN LHC reach at an integrated luminosity
 $\sim\ 100\ {\rm fb^{-1}}$ is indicated by the vertical orange dotted line: at 
neutralino masses larger than $m_{\tilde\chi_1}\simeq 500\ {\rm GeV}$ LHC will 
probably not be able to detect supersymmetry in the present parameter space 
setting.}\label{fig:FU50}
\end{figure*}

\subsection{$A$-Pole Funnels}\label{sec:funnel}
We now proceed in our analysis moving to the $A$-Pole Funnel regions. The defining
condition for this region is $2m_{\chi_1} \simeq m_A$, and again we show ($m_0,M_{1/2}$) points which saturate the limiting case for which equality holds.  In Fig.~\ref{fig:FU50} 
we consider  $\tan\beta=50$ and $\mu<0$. Also in this case, though the parameter space, 
as shown in Fig.~\ref{fig:PS50} is by far larger than  in the coannihilation 
strip, the spread in the LFV rates is again remarkably narrow, as can be inferred from
 the gray shaded region surrounding the $\brtm$ line. As before, the lower limit 
on $m_{\tilde\chi_1}$ is set by the upper bound on the inclusive branching ratio $b\rightarrow s \gamma$.
The projected LHC reach only extends up to 
$m_{\tilde\chi_1}\lesssim 500\ {\rm GeV}$, thus leaving a sizable portion of 
parameter space outside visibility at the future CERN facility. This is the
 first instance where LFV experiments  actually {\em compete} with CERN LHC 
as an \textit{additional} road to supersymmetry. In fact, values of $s_{13}\lesssim0.2$ are already ruled out by the current experimental bounds on $\brmu$. Should the experimental reach for this branching ratio be lowered down to $10^{-13}$, we
 would be able to  detect $\mu \to e \gamma$  signals in the whole funnel region, 
provided that $s_{13}\gtrsim{\mathcal O}(10^{-2})$. Interestingly, the large mass region lies beyond the expected LHC reach. As in the previous case, we also point out that an experimental sensitivity on $\brtm$ of $\mathcal{O}(10^{-8})$ would already start probing this region. 

\subsection{Focus Point Regions}\label{sec:focuspoint}
\begin{figure*}
\begin{center}
\includegraphics[scale=0.7]{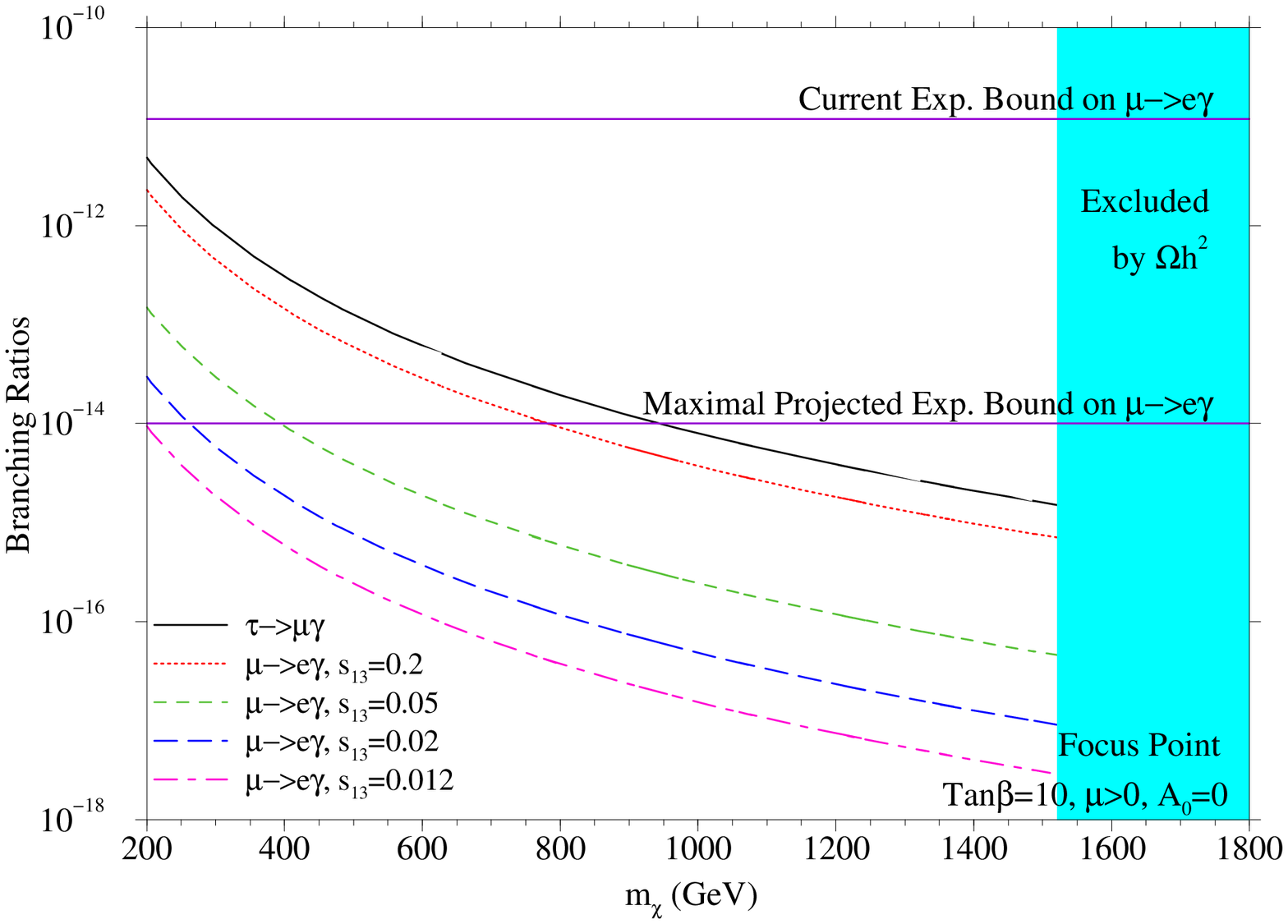}
\end{center}
\caption{\em $\brtm$and $\brmu$, for various values of 
$s_{13}=0.2,\ 0.05,\ 0.02$ and 0.012, along the extreme focus point region, in 
the $(M_{1/2},m_0)$ plane for $\tan\beta=50$, $\mu<0$ and $A_0=0$. The parameter 
space points we use here are those such that the higgsino content of the 
lightest neutralino is maximal. The cyan shaded region at large neutralino 
masses gives an $\Omega_{\tilde\chi_1} h^2$ exceeding the current WMAP constraint
on CDM density. We also show the current and projected sensitivities to $\brmu$.
The CERN LHC reach lies at neutralino masses {\em smaller} than 200 GeV. All the
SUSY parameter space points in this plot are therefore {\em outside} CERN LHC 
reach at an integrated luminosity $\sim\ 100\ {\rm fb^{-1}}$.}\label{fig:FOCUS10}
\end{figure*}

In the {\em focus point region} very large values of $m_0$ lower the Higgs
 mixing parameter $\mu$, thus entailing the generation of a non-negligible 
higgsino component in the lightest neutralino\footnote{It should be noted
that this situation takes naturally place in several other soft SUSY breaking scenarios, for instance with non-universal gaugino masses \cite{ref:gaugnonuniv}, where the lightest neutralino is mainly a higgsino;
relevant consequences for LFV can be found in Ref.~\cite{spcynonuniv}.}. 
This, in turn, yields an {\em enhancement} in the annihilation cross section 
with respect to the pure bino case, together with coannihilation effects 
with the next to lightest neutralino and, more importantly, with the lightest 
chargino, owing to the mass matrix structure of neutralinos and charginos. 
The combination of coannihilation effects and of a larger annihilation cross 
section forces the neutralino relic density to drop to very low values, which 
may be nonetheless compatible with the current dark matter abundance (see the 
discussion in Sec.~\ref{sec:darkmatter}).

The focus point region poses several computational problems, since it lies 
very close to parameter space points where EWSB fails, and moreover because 
it is rather fine-tuned, being very sensitive to the 
input parameters, especially the top mass, $m_t$. 
For this reason, this region of parameter space has 
sometimes not been included in CMSSM parameter space
 analyses \cite{Ellis:2002wv}. Nevertheless, we include it in 
our discussion. Indeed, we consider it useful to analyze the situation 
from the LFV rates point of view because it will be very hard
to probe most of this region at the LHC. 

As before, we once again resort to a low (10) and a large (50) value of 
$\tan\beta$, and we choose to show the extreme part of the focus point region, 
\ie\ that at the largest possible $m_0$, and hence where the higgsino content 
is maximal. Our choice is again motivated by two considerations: first, the region is 
sufficiently narrow so that LFV rates along the focus point region would appear in any case as single lines; second, choosing 
the maximal possible higgsino content we extend the parameter space line up to 
the largest possible neutralino masses.

The lower neutralino mass bound in the focus point region, due to the mass 
vicinity between the lightest neutralino and chargino, is typically dictated
 by the chargino mass bound from direct searches. As 
anticipated, the LHC reach is rather limited in this region: it lies around a neutralino mass of $\sim\ 200\ {\rm GeV}$. The large masses 
characterizing the sfermion spectrum naturally suppress LFV processes, but 
nonetheless, even in the less favorable case of $\tan\beta=10$, LFV can probe SUSY up to 
neutralino masses around 800 GeV, corresponding to very large values of the 
soft breaking masses at the GUT scale, namely $M_{1/2}\approx5.5\ {\rm TeV}$ 
and $m_0\approx17.5\ {\rm TeV}$. As regards the large $\tan\beta$ case, 
$\brmu$ may be within future experimental reach for neutralino masses 
{\em in the multi-TeV range}, provided $s_{13}$ is of ${\mathcal O} 
(10^{-1}\div10^{-2})$. Therefore, the focus point region, as well as, in 
general, the case of a higgsino dominated lightest neutralino, typically tends 
to {\em favor} $\brmu$ with respect to direct accelerator searches in the quest 
for supersymmetry\footnote{In a recent paper  \cite{Baer:2003ru}, it has been 
pointed out that at future $e^+\ e^-$ Linear Colliders with center of mass 
energy  $\sqrt{s}\approx 0.5\div 1\ {\rm TeV}$, the accelerator reach in the 
CMSSM focus point region may be by far larger than that at the CERN LHC}.

\begin{figure*}
\begin{center}
\includegraphics[scale=0.7]{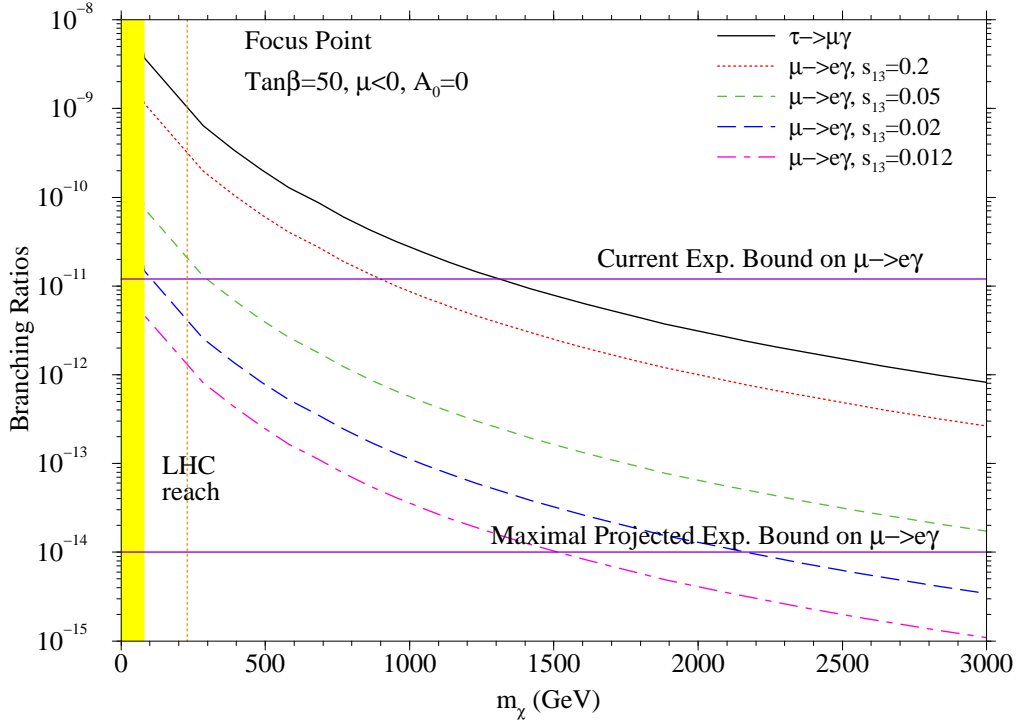}
\end{center}
\caption{\em $\brtm$and $\brmu$, for various values of 
$s_{13}=0.2,\ 0.05,\ 0.02$ and 0.012, along the extreme focus point region, 
in the $(M_{1/2},m_0)$ plane for $\tan\beta=50$, $\mu<0$ and $A_0=0$. Again, 
the parameter space points we use here are those such that the higgsino content 
of the lightest neutralino is maximal. We show neutralino masses up to 3 TeV, 
which are still allowed by relic density considerations. The expected sensitivity
of CERN LHC is showed by the vertical orange dotted line, while the yellow 
shaded area on the left indicates the bound stemming from the chargino mass 
limit set by LEP direct searches.}\label{fig:FOCUS50}
\end{figure*}

\begin{figure*}
\begin{center}
\includegraphics[scale=0.7]{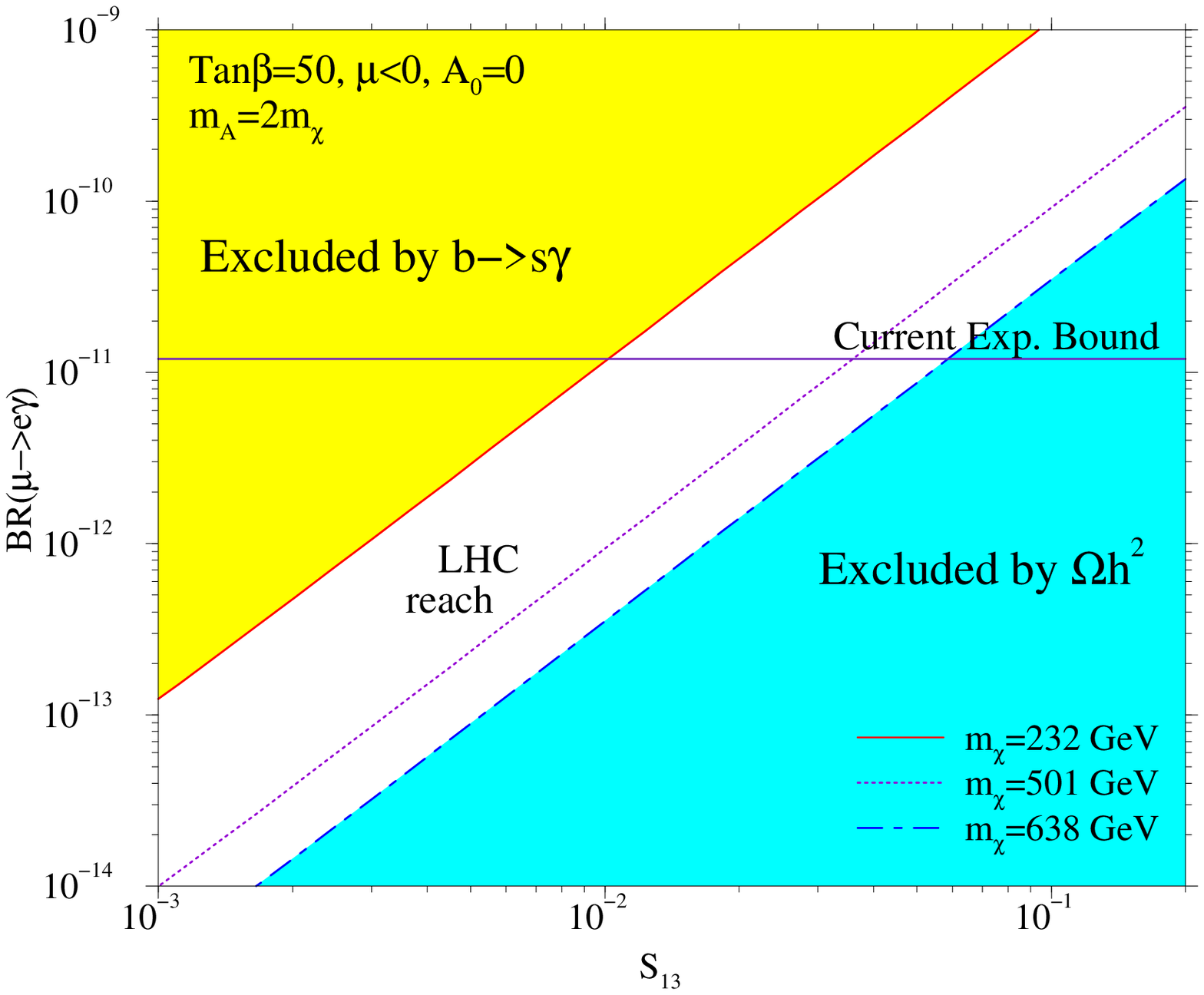}
\end{center}
\caption{\em The dependence of $\brmu$ on $s_{13}$, for $\tan\beta=50$, 
$\mu<0$ and $A_0=0$ along the parameter space line, in the $(M_{1/2},m_0)$ 
plane corresponding to $2\cdot m_{\tilde\chi_1}=m_A$, \ie\  in the central part 
of the funnel region. The upper region, shaded in yellow, is ruled out by the
 $b\rightarrow s\gamma$ bound, while the lower region, shaded in cyan, is 
disallowed by the $\Omega_{\tilde\chi_1} h^2$ bound on the neutralino relic 
density. The three lines respectively correspond to the lower and the upper 
neutralino mass limits and to the largest neutralino mass within LHC reach. 
We also report the current experimental upper bound on $\brmu$.}\label{fig:S13}
\end{figure*}
\section{The role of $U_{e3}$}
The importance of the parameter $U_{e3}$ in its connection with the neutrino mixing 
matrix and lepton flavour violation in the SUSY
seesaw is apparent from the discussions we carried out in the previous
two sections.
The precise value of $U_{e3}$ turns out to be of critical importance in at least two contexts in the present analysis:\\

\noindent 
(i) \textbf{LFV versus LHC} \\
Within the coannihilation
 regions, as we have seen, LFV will play only a ``supporting'' r\^ole to the more powerful LHC searches. 
The first instance where LFV may reveal itself as a superior tool arises in the heavy mass A-pole
funnel region. However, the prominence of LFV searches is crucially dependent
on $U_{e3}$. To make this more precise, we plot in Fig.~\ref{fig:S13} the $\brmu$ as a function of $s_{13}$, showing iso-neutralino mass curves. 
From the plot it is evident that an experimental sensitivity of $10^{-13}$ would
 allow to detect LFV as long as $s_{13}\gtrsim 5\cdot 10^{-3}$. On the other hand, 
there is a large band, at heavy SUSY particles masses, lying beyond LHC reach, 
which will be fully accessible to LFV experiments. The same holds true in the focus point region, where, provided $U_{e3}$ is not too small, $\brmu$ will probe SUSY far more effectively than the LHC.\\

\begin{figure*}
\begin{center}
\includegraphics[scale=0.65]{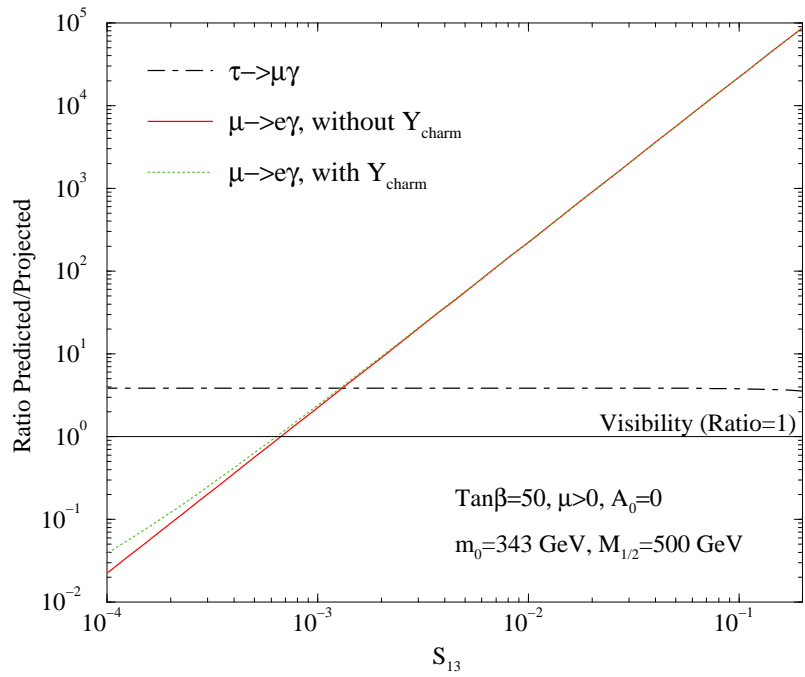}
\end{center}
\caption{\em The ratio of the predicted $\brtm$ $($respectively $\brmu)$ and 
the approximate projected maximal sensitivity of $10^{-8}$ $($resp. $10^{-13})$ 
at a particular parameter space point along the coannihilation strip at 
$\tan\beta=50$, vanishing $A_0$ and positive $\mu$. The green dotted line 
corresponds to the case where the effect of a second non-zero neutralino Yukawa 
coupling, set equal to the charm quark Yukawa coupling, is taken into 
account.}\label{fig:COMP}
\end{figure*}

\noindent 
(ii) \textbf{$\mu \to e\gamma$ versus $\tau \to \mu \gamma$} \\
We have seen that, if $s_{13}$ is not too small, the constraints coming 
from $\mue$ are stronger than those derived from $\tmu$.  Hence, there should 
be a {\em critical value} of $s_{13}$ below which $\tmu$ becomes more relevant than 
$\mue$. What is exactly this value? To answer this question we show in 
Fig.~\ref{fig:COMP} $\brmu$ in units of $10^{-13}$ as a function of $s_{13}$,
for the particular point at $m_0=343$, $M_{1/2}=500$, $\tan \beta$ = 50 and $\mu > 0$.
The dashed-dotted line denotes the value of $\brtm$ in units of $10^{-8}$.
We observe that the intersection between the $\tmu$ line -- which is almost independent of $s_{13}$ -- and the $\mue$ line takes place at a 
value of $s_{13}$ close to $10^{-3}$. If $s_{13}\gtrsim 10^{-3}$ then $\mue$ is more likely to be 
observed in future experiments than $\tmu$. Although these results were obtained 
for a specific value of $m_0$ and $M_{1/2}$, changes in these parameters are 
expected to affect equally both processes so that the critical value 
$s_{13}\approx 10^{-3}$ does not depend on $m_0$ or $M_{1/2}$ (We have also 
checked numerically that this is indeed the case). 
When $s_{13}$ is very small, the contribution to $\brmu$, proportional to 
the second Yukawa coupling dominates over that proportional to the top Yukawa 
coupling. In same figure, we show the predictions for $\brmu$ \emph{with} and 
\emph{without} taking into account the effect of the second Yukawa coupling, which we
set equal to the charm Yukawa \cite{oscar}.  
Notice that such effect is only relevant for $s_{13}\lesssim 7 \times 10^{-4}$.   

\section{Summary of the Results}
Here we summarise our results on the complementarity
of the three search roads for the three allowed regions of the CMSSMRN in
the `best case' scenario. 
\begin{itemize}
\item \textit{Coannihilation Regions}: In these regions, which are
mostly accessible at LHC, an improvement of two orders of magnitude in the branching
ratio sensitivity would make $\mu \to e \gamma$ visible for most
of the parameter space as long as $s_{13} \gtrsim 0.02$, even for the 
low tan $\beta$ region. For large tan $\beta$, independent of $s_{13}$,
$\tau \to \mu \gamma$ will start probing this region provided a sensitivity
of $\mathcal{O}(10^{-8})$ is reached.  
\item \textit{A-pole funnel Regions}: In these regions the LHC
reach is not complete and LFV may be competitive. If $s_{13} \gtrsim 10^{-2}$,
the future $\mu \to e \gamma$ experiments will probe most of the parameter space regions. 
As before, $\tau \to \mu \gamma$ will probe this region once the BR sensitivity reaches
 $\mathcal{O}(10^{-8})$.
\item \textit{Focus Point Regions}: Since the LHC reach in this region is rather limited due to the
large $m_0$ and $M_{1/2}$ values, LFV could constitute a privileged road towards SUSY discovery. As we pointed out in Sec.~\ref{sec:model}, DM searches will also have in future partial access to this region, leading to a new complementarity between LFV and the quest for the cold dark matter constituent of the universe.
\end{itemize}

\section{Outlook}\label{sec:conclusions}
So far, the three roads towards SUSY we mentioned are (a) direct accelerator SUSY particles searches; (b) indirect searches through rare FCNC/CP violating processes and (c) direct and indirect DM searches. In this 
decade, these three ways are going to remain the best tools to get 
evidence for SUSY with foreseeable impressive improvements on their
sensitivity reach (in particular, for (a), with the advent of LHC). Our work 
tries to provide a critical assessment on their complementarity, focusing
on the maximal reach of each of them in various viable parameter space regions. To make such an analysis quantitative,
one needs a specific low-energy realisation. We chose to work in the
context of the CMSSMRN model, because (i) `CMSSM' is the prototype of `safe'
flavour blind low-energy SUSY extensions to the SM compatible with all the precision SM tests 
and (ii) `RN' provides an appealing mechanism to generate neutrino masses. This model
is severely constrained by direct SUSY particles searches, FCNC constraints, and,
even more strongly, by the information we gained about the amount of CDM from WMAP and
large scale structure analysis. As a matter of fact, only three regions of the CMSSMRN parameter
space still survive. In the most favorable (`best') case, LFV is certainly complementary to LHC searches in all these three regions and, interestingly, it can probe portions of the parameter space which will be unaccessible to the LHC. 

\section*{Acknowledgments}
AM and SV acknowledge support from the RTN European 
project ``Physics Across the Present Energy Frontier'' HPRN-CT-2000-0148 and from the "Italian
University and Research Ministry" under the program "PRIN:Astroparticle Physics" 2002. 


\end{document}